\documentclass{aa}

\usepackage{graphicx}
\usepackage{natbib}

\newcommand{\subfigimg}[3][,]{%
  \setbox1=\hbox{\includegraphics[#1]{#3}}
  \leavevmode\rlap{\usebox1}
  \rlap{\hspace*{10pt}\raisebox{\dimexpr\ht1-2\baselineskip}{#2}}
  \phantom{\usebox1}
}

\usepackage{amsmath}    
\usepackage{amssymb}    

\begin{document}

\title{The Galactic habitable zone around M and FGK stars with chemical evolution models with dust}

\author {E. Spitoni\inst{1} \thanks {email to:
    spitoni@oats.inaf.it} \and L. Gioannini\inst{1} \and F. Matteucci\inst{1,2}} \institute{
  Dipartimento di Fisica, Sezione di Astronomia,  Universit\`a di Trieste, via
  G.B. Tiepolo 11, I-34131, Trieste, Italy \and I.N.A.F. Osservatorio
  Astronomico di Trieste, via G.B. Tiepolo 11, I-34131, Trieste,
  Italy}

\date{Received xxxx / Accepted xxxx}

\abstract {The Galactic habitable zone is defined as the region with
  highly enough metallicity to form planetary systems in which
  Earth-like planets could be born and might be capable of sustaining
  life surviving to the destructive effects of nearby supernova
  explosion events. }{Galactic chemical evolution models can be useful
  tools for studying the galactic habitable zones in different
  systems.  Our aim here is to find the Galactic habitable zone using
  chemical evolution models for the Milky Way disc, adopting the most
  recent prescriptions for the evolution of dust and for the
  probability of finding planetary systems around M and FGK
  stars. Moreover, for the first time, we will express those
  probabilities in terms of the dust-to-gas ratio of the ISM in the
  solar neighborhood as computed by detailed chemical evolution
  models.}{At a fixed Galactic time and Galactocentric distance we
  determine the number of M and FGK stars having Earths (but no gas
  giant planets) which survived supernova explosions, using the  formalism
  of our Paper I.} {The probabilities of finding
  terrestrial planets but not gas giant planets around M stars deviate
  substantially from the ones around FGK stars for supersolar values
  of [Fe/H]. For both FGK and M stars the maximum number of stars
  hosting habitable planets is at 8 kpc from the Galactic Centre, if
  destructive effects by supernova explosions are taken into account.
  At the present time the total number of M stars with habitable
  planets are $\simeq$ 10 times the number of FGK stars. Moreover, we
  provide a sixth order polynomial fit (and a linear one but more
  approximated) for the relation found with chemical evolution models
  in the solar neighborhood between the [Fe/H] abundances and the
  dust-to-gas ratio.  }{The most likely Galactic zone to find
  terrestrial habitable planets around M and FGK stars is the annular
  region 2 kpc wide centred at 8 kpc from the Galactic center (the
  solar neighborhood). We also provide the probabilities of finding
  Earth-like planets as the function of the ISM dust-to-gas ratio
  using detailed chemical evolution models results. }


\titlerunning{The Galactic habitable zone in presence of dust }
\authorrunning{Spitoni et al.}
\maketitle

\section{Introduction}
The Galactic habitable zone (GHZ) has been defined as the region with
sufficiently high abundances of heavy elements to form planetary
systems in which terrestrial planets could be found and might be
capable of sustaining life.  Therefore, the chemical evolution of the
Galaxy plays a key role to properly model the
GHZ evolution in space and in time. 
 The minimum metallicity needed for planetary formation, which
would include the formation of a planet with Earth-like
characteristics (firstly discussed by Gonzalez et al. 2001) has been
fixed at the value of 0.1 Z$_{\odot}$ by the theoretical work of
Johnson \& Li 2012).

In the last years several
purely chemical evolution models (Lineweaver 2001, Lineweaver et al. 2004,
 Prantzos 2008,
Carigi et al 2013, Spitoni et al. 2014) have studied the  habitable zones
of our Galaxy as functions of the Galactic time and Galactocentric distances.
In particular Spitoni et al. (2014, hereafter Paper I),  in which radial gas flows were 
included,  confirmed the previous results of Lineweaver at al. (2004) and 
found that the maximum number of stars which can host habitable
terrestrial planets are in solar neighborhood, i.e. the region centered
at 8 kpc, and 2 kpc wide.

Recently, the GHZ has been also studied in the cosmological context
($\Lambda$CDM) by Forgan et al. (2015), Gobat \& Hong (2016), 
Zackrisson et al. (2016), and Vukoti{\'c} et al. (2016) showing which
kind of halos can give rise to galactic structures in which habitable
planets could be formed.

In  most of the models mentioned above (both for purely chemical evolution and
cosmological models), it was considered the probability of forming
planetary systems in which terrestrial planets are found without
any gas giant planets or hot Jupiters, because in principle the last
two planet types could destroy Earths during their evolution.

The terms giant planets, gas giants, or simply Jupiters, refer to
large planets, typically $>$10 M$_{\oplus}$, that are not composed
primarily of rock or other solid matter. When orbiting close to the
host star they are referred to as hot Jupiters or very hot Jupiters.
The ``core accretion model'' of giant planet formation is the most
widely accepted in the literature, and the next stochastic migration
due to turbulent fluctuations in the disc could destroy the
terrestrial planets.  Rice \& Armitage (2003) showed that the
formation of Jupiter can be accelerated by almost an order of
magnitude if the growing core executes a random walk with an amplitude
of $\approx$ 0.5 AU.  Nowadays, about 3 thousands of planets have been
discovered and the statistics is   good enough to confirm that most
of planetary systems host planets which are not present in our solar
system, such as hot Jupiters or super-Earths.

 In this paper we retain the assumption that gas giant planets
could destroy, during their evolution, terrestrial ones (we are aware
that the real effects are still uncertain).  On this basis we study
the GHZ using the most updated probabilities related to the formation
of gas giant planets as functions of [Fe/H] abundance values as well
as the stellar mass for FGK and M stars.  It is recent the discovery
of a planetary system around the M star Trappist-1 (Gillon at
al. 2016, 2017) composed by seven terrestrial planets characterized by
equilibrium temperature low enough to make possible the presence of
liquid water. This detection makes the habitability around M stars
even more interesting.

  In this paper  we  also compute the probabilities related to the formation of gas giant planets as
functions of [Fe/H] abundance values and the stellar mass for 
FGK and M stars with a detailed chemical evolution
model for the Milky Way  with dust evolution.

Even though, we know very little about the formation of planetary
systems: in particular, it is not well understood the transition from
a protoplanetary disc to a planetary system.  In this transition, dust
and gas rapidly evolve in very different ways due to many processes
(Armitage 2013) such as dust growth, gas photoevaporization (Alexander
et al. 2014), gas accretion onto the star (Gammie 1996).  Dust plays a
fundamental role in the formation of the first planetesimals, as it
represents the solid compounds of the matter which can form rocky
planetesimals and therefore planets.

The first fundamental step to understand and explain the origin of the
observed diversity of exoplanetary systems, is to measure the
stellar disc properties, especially the disc mass.  Dust grains of
$\mu$m are directly observed in protoplanetary discs and then, dust
coagulation increase their size up to mm (Dullemond $\&$ Dominik
2005). On the other hand, observations of the gas are almost forbidden, 
because  it is relatively cool and in molecular form.

 In most cases, the mass of the gas is set starting from the one of
 the dust and by assuming a value for the dust-to-gas
 ratio. Unfortunately, this practice has several uncertainties
 (Williams \& Best 2014).  Usually, models of planetary formation use
 an average value of $10^{-2}$ for the initial condition of the
 dust-to-gas ratio in the protoplanetary disc (Bohlin et al. 1978).

 Furthermore, the formation rate of gas giant planets seems to be
related to the metallicity of the hosting stars (Fisher \& Valenti
2005, Johnson et al. 2010, Mortier et al. 2012, Gaidos \& Mann 2014).
Moreover, even if  Buhhave et al. (2012) showed that planets with radii
smaller than 4 R$_{\oplus}$ do not  present any metallicity
correlation, the theoretical work of Johnson \& Li (2012) predicted that
first Earth-like planets likely formed from
circumstellar discs with metallicities  Z $\geq$ 0.1 Z$_{\odot}$.

 With this work, providing the time evolution of the
dust, we discuss the connection between the metallicity of the ISM and
the dust-to-gas ($\frac{D}{G}$) ratio (especially for the solar
neighborhood).  In this way, we can link the metallicity of stars,
which is observationally related to the probability of the presence of
hosted planets, with the initial dust-to-gas ratio of the
protoplanetary discs (the dust-to-gas ratio of the ISM at the instant
of the protoplanetary disc formation).

The paper is organized as follows: in Section 2 we present the
probabilities of terrestrial planets around M and FGK stars.
 In Section
3  we describe the  Milky Way chemical evolution model  with
dust and we present  the main results  in Section 4.  Finally, our
conclusions are summarized in Section 5.
 
\begin{figure*}
\centering
\begin{tabular}{@{}p{0.45\linewidth}@{\quad}p{0.45\linewidth}@{}}
\subfigimg[width=\linewidth]{ \hspace{ 0.6cm}  A)}{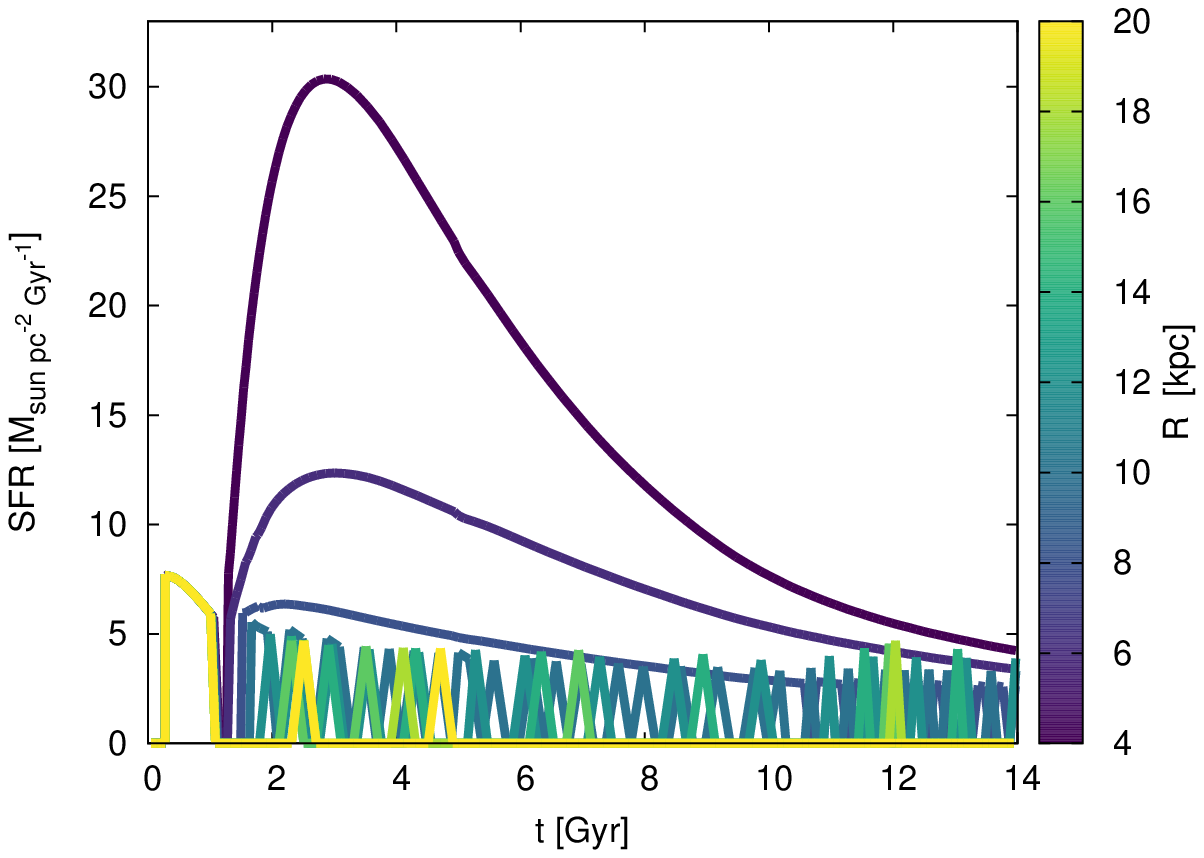} &
    \subfigimg[width=\linewidth]{ \hspace{ 0.6cm} B)}{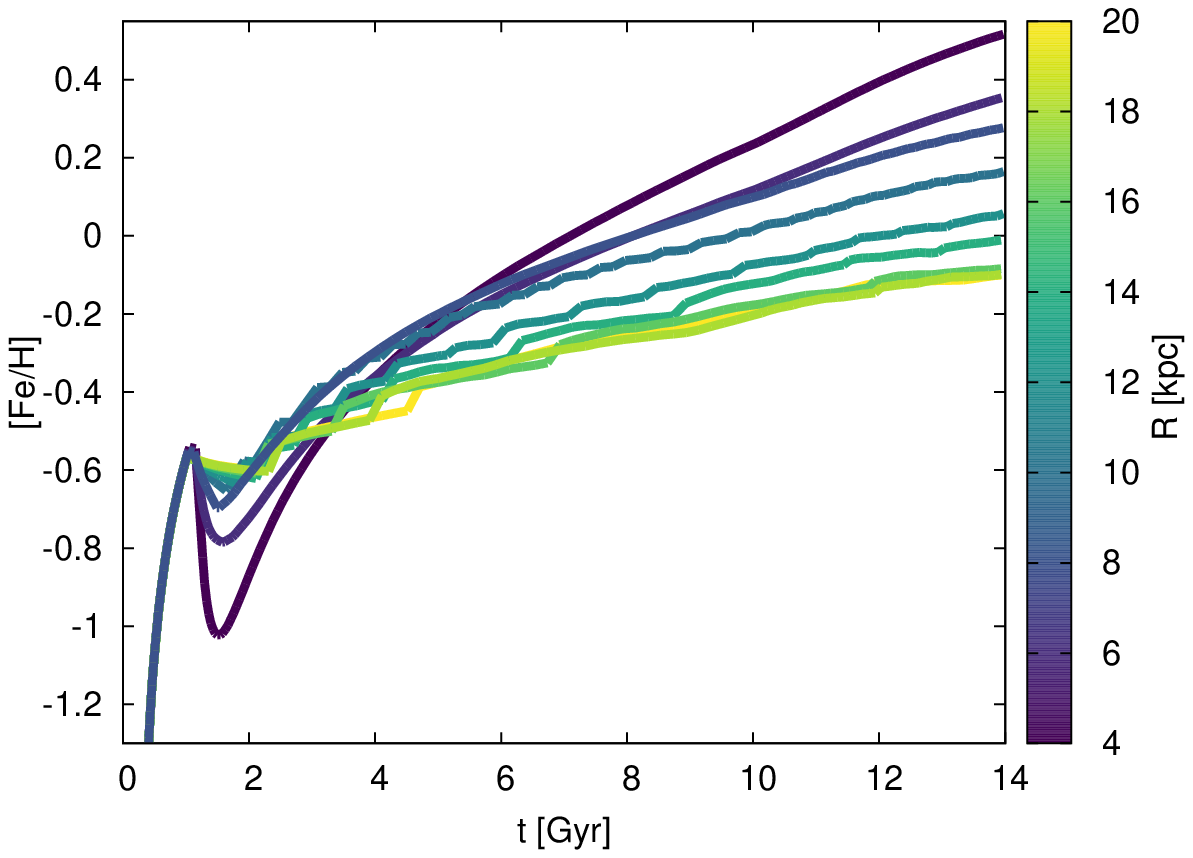} \\
    \subfigimg[width=\linewidth]{\hspace{ 0.75cm} C)}{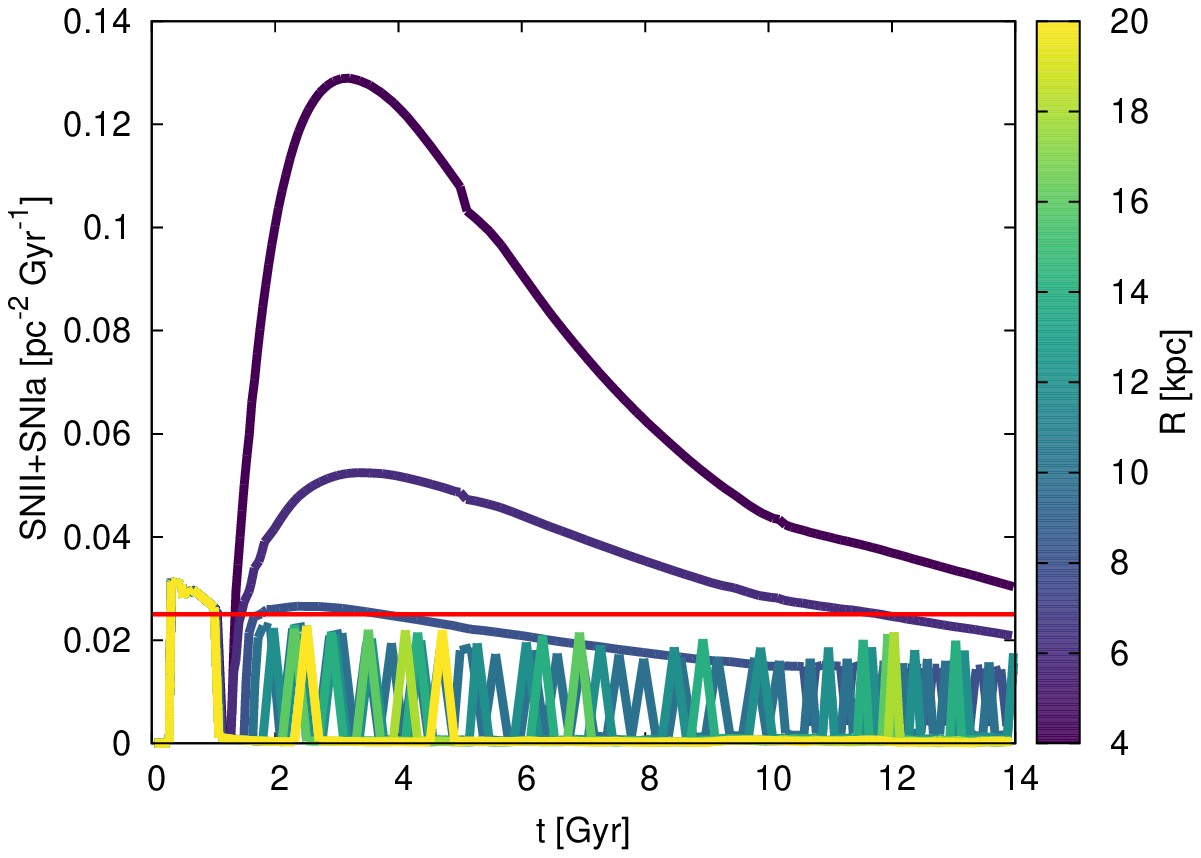} &
    \subfigimg[width=\linewidth]{\hspace{ 0.75cm} D)}{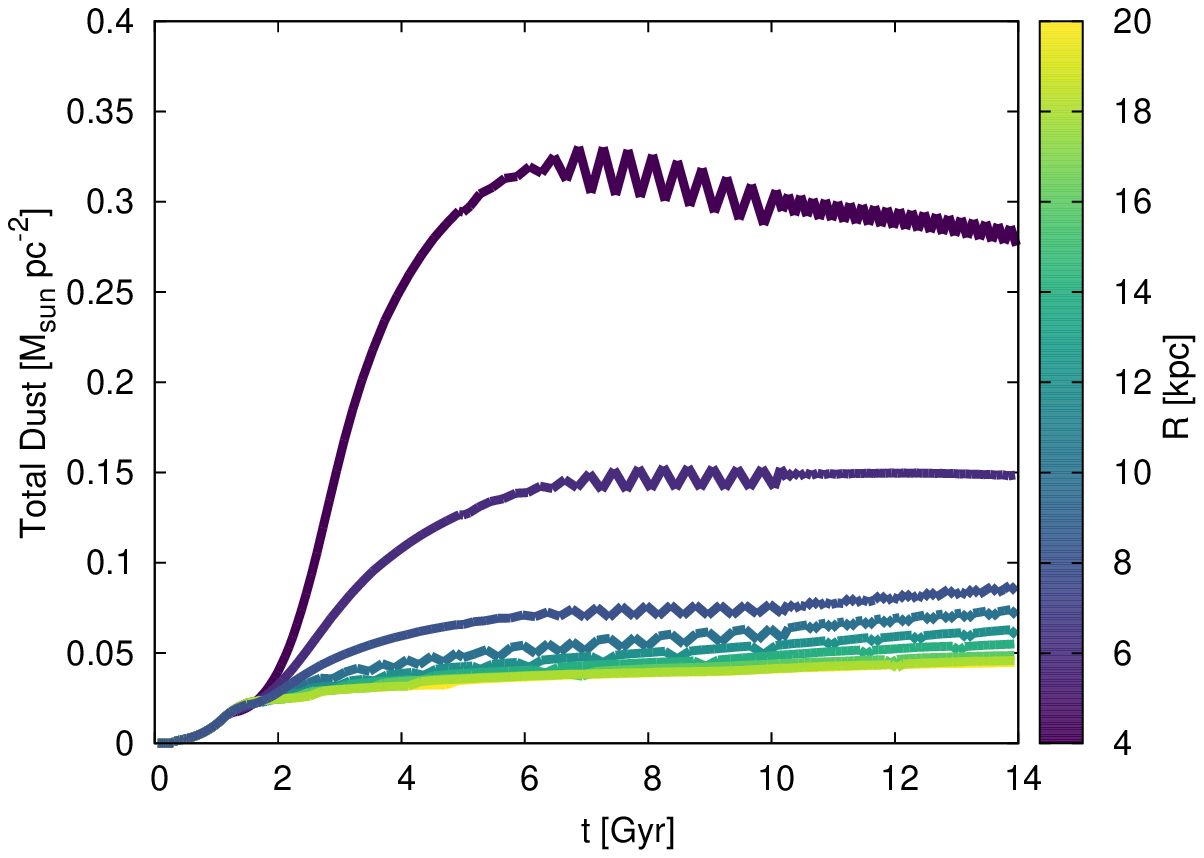}
  \end{tabular}
  \caption{{\it Panel} A): The SFR as the function of the Galactic
    time; {\it Panel} B): The time evolution of the [Fe/H] abundances
    with the two-infall chemical evolution model for the Milky
    Way disc (the ``age-metallicity'' relation); {\it Panel} C): The
    evolution in time of the Type II SN rates plus the Type Ia SN
    rates. With the red line we label the quantity $2 \times
    <RSN_{SV}>$ which represents the minimum SN rate value  (adopted in this work
    and in Paper I) to have  destruction effects from
    SN explosions; {\it Panel} D): The time evolution of the total
    dust surface mass density. The color code in the four panels indicates
    the different Galactocentric distances.} 
 \label{SFR}
\end{figure*}

\section{The probabilities of terrestrial planets around M and FGK stars}

Buchhave et al. (2012) who analyzed the mission Kepler, found that
  the frequencies of the planets with earth-like sizes are almost
  independent of the metallicity of the host star up to [Fe/H]
  abundance values smaller than 0.5 dex.

 In agreement with these observations, Prantzos (2008) fixed the
 probability of forming Earth-like planets ($P_{FE}$, where $FE$
 stands for Forming Earths) at value of 0.4 for [Fe/H] $\geq$ -1 dex,
 otherwise $P_{FE}=0$ for smaller values of [Fe/H].  This assumption
 was also adopted by Carigi et al. (2013) and Paper I. The
 value of $P_{FE}$= 0.4 was chosen to reproduce the metallicity
 integrated probability of Lineweaver et al. (2001).

In Paper I it was considered the case in which gaseous
 giant planets with same host star can destroy terrestrial planets
 (i.e. during their migration path). 
 Armitage (2003) pointed out the potentially hazardous effects of the gas giant
planet migration on the formation of Earth-like ones, and suggested that
these planets preferentially exist in systems where massive giants did
not migrate significantly.  Matsumura et al. (2013) studying the
orbital evolution of terrestrial planets when gas giant planets become
dynamically unstable, showed that Earth-like planets far away from
giants can also be removed. 

 On the other hand, various numerical simulations found that the
formation of earths is not necessarily prevented by the gas giant
planet migration, when eccentricity excitation timescales for
(proto-)terrestrial planets are long compared to migration timescales
of giant planets (e.g., Mandell \& Sigurdsson 2003; Lufkin et
al. 2006; Raymond et al. 2006).

 The adopted probability of
 formation of a gaseous giant planet as the function of the iron
 abundance in the host star is taken by Fischer \& Valenti (2005) as:
\begin{equation}
 P_{GGP}\left(\mbox{[Fe/H}]\right)= 0.03 \times 10^{2.0 \mbox{[Fe/H]}}.
\label{valenti}
\end{equation}

A possible theoretical explanation is that the high metallicity observed in
some stars hosting giant planets represent the original composition that
protostellar and protoplanetary molecular clouds were formed.  In this
scenario, the higher the  metallicity of the primordial cloud, the
proportion  of dust to gas in the protoplanetary disc. This
facilitates the condensation and accelerates the protoplanetary
accretion before the disc gas is lost (Pollack et al. 1996). Giant
planets are subsequently formed
by runaway accretion of gas onto
such rocky cores with M $\approx$ 10$M_{\oplus}$, rather than by gravitational instabilities in a gaseous disk which predicts formation much less sensitive to metallicity (Boss, 2002).

The novelty of this work is to consider the following new probabilities for gas giant planets formation found by Gaidos \& Mann (2014)  and used by Zackrisson et al. (2016)  around FGK and M stars
which are functions also of the   masses of the hosting stars:

\begin{equation}
 P_{GGP/FGK}\left(\mbox{[Fe/H}], M_{\star}\right)= 0.07 \times 10^{1.8 \mbox{[Fe/H]}} \left( \frac{M_{\star}}{M_{\odot}}\right),
\label{eqgiantF}
\end{equation}
for the FGK stars, and

\begin{equation}
 P_{GGP/M}\left(\mbox{[Fe/H}], M_{\star}\right)= 0.07 \times 10^{1.06  \mbox{[Fe/H]}} \left( \frac{M_{\star}}{M_{\odot}}\right),
\label{eqgiantM}
\end{equation}
and for M stars, where $M_{\star}$ is the mass of the host star in units of solar masses.

We assume that the range of masses spanned by $M$ type stars is the
following one: $0.08 \leq \frac{M_{\star}}{M_{\odot}} \leq 0.45$. For
the $FGK$ the range is: $0.45 \leq \frac{M_{\star}}{M_{\odot}} \leq
1.40$.

The probability of forming terrestrial planets around FGK/M stars but not gaseous giant planets is given by:

\begin{equation}
P_{E/FGK,M}=P_{FE} \times (1-P_{GGP/FGK,M}).
\label{ter}
\end{equation}
Here, we make the conservative assumption of Prantzos (2008) and
Paper I that the $P_{FE}$ probability is constant at the
value of 0.4 for all stellar types, including  M and FGK stars. 
 On the other hand, Zackrisson et al. (2016) presented
results where $P_{FE}=0.4$ around FGK stars, and $P_{FE}=1$ around M
stars.

 The
possibility of finding habitable planets around M-dwarf has long been
debated, due to differences between the unique stellar and planetary
environments around these low-mass stars, as compared to hotter, more
luminous Sun-like stars (Shields et al. 2016). 
  The
presence of multiple rocky planets (Howard et al. 2012), with roughly
a third of these rocky M-dwarf planets orbiting within the habitable
zone, supports the hypothesis of the presence of liquid water on their
surfaces. 
 On the other hand, flare activity, synchronous rotation,
and the likelihood of photosynthesis could have a severe inpact on the
habitability of planets hosted by M dwarf stars (Tarter et al., 2007).

We define $P_{GHZ}(FGK/M,R,t)$ as the fraction of all  FGK/M  stars
having around Earths (but no gas giant planets) which survived supernova
explosions as a function of the Galactic radius and time:
\begin{displaymath}
P_{GHZ}(FGK/M,R,t)= 
\end{displaymath}
\begin{equation}
= \frac{\int_0^t SFR(R,t') P_{E/FGK,M} (R,t') P_{SN}(R,t') dt'}{\int_0^t SFR(R,t')dt'.}
\label{GHZ}
\end{equation}

This quantity must be interpreted as the relative
probability to have complex life around one star at a given position,
as suggested by Prantzos (2008).

In eq. (\ref{GHZ}) $SFR(R,t')$ is the star formation rate (SFR) at the
time $t'$ and Galactocentric distance $R$, and $P_{SN}(R,t')$ is the probability of
surviving supernova explosion.

We know that hard radiation originated by close-by SN explosions could
lead to the depletion of the ozone layer in terrestrial atmosphere. At
this point, the  ultraviolet radiation from the host star can penetrate the
atmosphere, altering and damaging the DNA and eventually causing the total
sterilization of the planet (Gehrels et al. 2003).

For this quantity we refer to the ``case 2'' model  of Paper I
in which the SN
destruction is effective if the SN rate at any time and at any radius
has been higher than twice the average SN rate in the solar neighborhood
during the last 4.5 Gyr of the Milky Way life (we call it
$<RSN_{SV}>$).

Therefore, we impose that if SN rate is larger than $2 \times <RSN_{SV}>$
  then $P_{SN}(R,t)$ = 0 else $P_{SN}(R,t)$ = 1. 
We also show results when SN effects are not taken into account, in
this case we simply impose $P_{SN}(R,t)$ = 1 at any time and galactic radius.
The ``case 2'' condition is almost
the same as that used by Carigi et al. (2013) to describe
their best models,  motivated by the fact that  life on Earth
has proven to be highly resistant, and 
the real effects of SN explosions on life are still extremely uncertain .

For $<RSN_{SV}>$ we adopt the value of 0.01356 Gyr$^{-1}$ pc$^{-2}$
using the results of the S2IT model of Spitoni \& Matteucci (2011) and Paper I.

Finally, we define the total number of stars formed at a certain time
$t$ and \ Galactocentric distance $R$ hosting Earth-like planet with
life $N_{\star \, life}(FGK/M,R,t)$, as: 

\begin{equation}
N_{\star \, life}=P_{GHZ} \times N_{\star tot} ,
\label{star}
\end{equation}
where  $N_{\star tot}(FGK/M,R,t)$ is the total number of stars created up to  time $t\
$ at the Galactocentric distance $R$.

\section{The Milky Way chemical evolution model with dust}
 To trace the chemical evolution of the Milky Way we adopt an updated 
version of the two-infall model of    Paper I in which we consider
the dust evolution using the new prescriptions of Gioannini et al. (2017).

\subsection{The two infall model of Paper I}
The chemical evolution model of Paper I is based on the
classical two-infall model of Chiappini et al. (2001). We describe
here the main characteristics of the model.

 We define $G_{i}(t)=G(t)X_{i}(t)$ as the fractional mass of the
 element $i$ at the time $t$ in the ISM, where $X_i(t)$ represents the
 abundance of the element $i$ in the ISM at the time $t$. The temporal
 evolution of $G_{i}(t)$ in the ISM is described by the following
 expression:
\begin{equation} \label{basic_model}
\dot {G}_{i}(t) = -\psi(t)X_i(t) + R_i(t) + \dot {G}_{i,inf}(t).
\end{equation}
 
The first term in the right side of eq. (\ref{basic_model}) represents
the rate at which the fraction of the element $i$ is subtracted by the
ISM due to the SFR process.  $R_i(t)$ is the returned mass fraction of
the element $i$ injected into the ISM from stars thanks to stellar
winds and SN explosions. This term takes into account nucleosynthesis
prescriptions concerning stellar yields and supernova progenitor
models.
The third term of eq.(\ref{basic_model}) represents the rate of the
infall of the element $i$. The infalling gas is not pre-enriched and
has a pure primordial composition.

  The two-infall approach is a sequential   model in  which the
  halo-thick disc and the thin disc form  by means of two independent infall
  episodes of primordial gas following this infall rate law:

 \begin{equation}            
 \dot {G}_{i,inf}(t) = a(r) e^{-t/ \tau_{H}}+ b(r) e^{-(t-t_{max})/ \tau_{D}(r)},
\end{equation}
where $\tau_{H}$  is  the typical timescale for the formation of the halo and thick disc and it is fixed to the value of
0.8 Gyr, while $t_{max}$ = 1 Gyr is the time for the
maximum infall onto the thin disc.
The coefficients $a(r)$ and $b(r)$ are obtained by imposing a
fit to the observed current total surface mass density in the thin
disc as a function of Galactocentric distance given by:
\begin{equation}
\Sigma(r) = \Sigma_0 e^{-R/R_D},
\end{equation}
where $\Sigma_0$ = 531 M$_{\odot}$ pc$^{-2}$ is the central total
surface mass density and $R_D$ = 3.5 kpc is the scale length.
\begin{figure}
\includegraphics[scale=0.7]{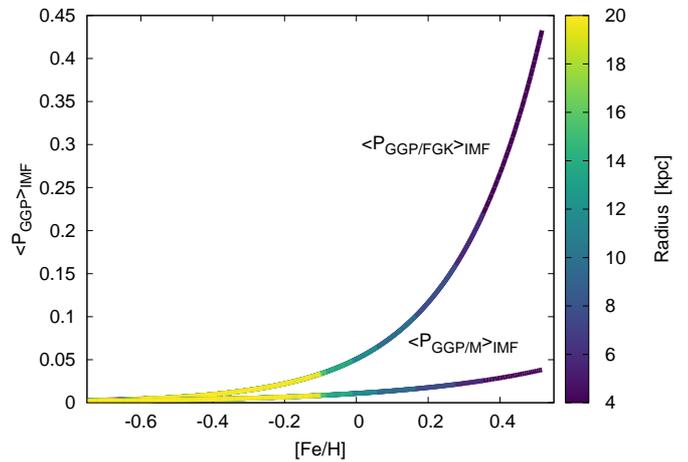}
  \caption{The probabilities $<$$P_{GGP/FGK}$$>$$_{IMF}$ and $<$$P_{GGP/M}$$>$$_{IMF}$  to find gas giant planets around FGK and M
 stars, respectively as functions of the abundance ratio [Fe/H]
 using our chemical evolution model for the Milky Way disc and
 adopting the prescriptions given by Gaidos \& Mann (2014). The color code 
indicates the Galactocentric distance. }  
\label{giant}
\end{figure} 
\begin{figure}
\includegraphics[scale=0.7]{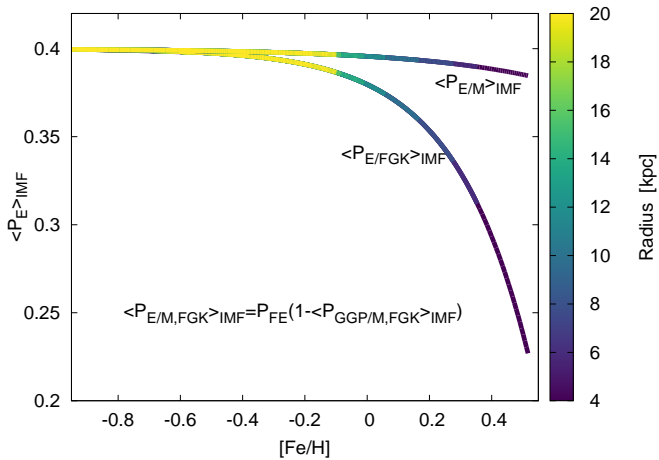}
  \caption{The probabilities $<P_{E/FGK}>_{IMF}$  and $<P_{E/M}>_{IMF}$ to find terrestrial planets but not gas
 giant planets around FGK and M stars, respectively, as functions of
 the abundance ratio [Fe/H] using our chemical evolution model
 for the Milky Way disc and adopting the prescriptions given by Gaidos
 \& Mann (2014) and Paper I. The color code indicates the Galactocentric
 distance. } \label{pe}
\end{figure} 
 Moreover, the
formation timescale  of the thin disc $\tau_{D}(r)$ is assumed to be a function of
the Galactocentric distance, leading to an inside-out scenario for the
Galaxy disc build-up. 
In particular, we assume that:
\begin{equation}
\tau_{D}(r)= 1.033 \, R\,  \mbox{(kpc)} - 1.267 \, \mbox{Gyr}.
\end{equation}
The Galactic thin disc is approximated by
several independent rings, 2 kpc wide, without exchange of matter
between them. 
A threshold gas density of 7 M$_{\odot}$ pc$^{-2}$ in the SF process
(Kennicutt 1989, 1998; Martin \& Kennicutt 2001; Schaye 2004) is also
adopted for the disc. The halo has a constant surface mass density as
a function of the Galactocentric distance at the present time equal to
17 M$_{\odot}$ pc$^{-2}$ and a threshold for the star formation in
the halo phase of 4 M$_{\odot}$ pc$^{-2}$, as assumed for the model B
of Chiappini et al. (2001).

 The assumed IMF is the one of Scalo (1986), which is assumed  constant in
  time and space. 

 The adopted law for the SFR is
  a Schmidt (1959) like one:
\begin{equation}
\Psi \propto \nu \Sigma^k_{gas}(r,t),
\label{schmidt}
\end{equation}

here $\Sigma_{gas}(r,t)$ is the surface gas density with the exponent
$k$ equal to 1.5 (see Kennicutt 1998; and Chiappini et al. 1997). The
quantity $\nu$ is the efficiency of the star formation process, and 
is constant and fixed to be equal to 1 Gyr $^{-1}$. 
The chemical  abundances are normalized to the Asplund et al. (2009) solar values. 
\subsection{Evolution of dust}
Our chemical evolution model also traces the dust evolution in the interstellar medium (ISM).
Defining $G_{i,dust}(t)$, we can thus write the equation for dust evolution as follows:
 
\begin{equation} \label{chem-DUST}
\begin{split}
\dot {G}_{i,dust}(t)& = -\psi(t)X_{i,dust}(t) + \delta_i R_i(t)  + \left( \dfrac{{G}_{i,dust}(t)}{\tau_{accr}}\right) \\ 
& \quad - \left( \dfrac{{G}_{i,dust}(t)}{\tau_{destr}} \right). \\
\end{split}
\end{equation}

The right hand of this equation contains all the processes which
govern the so called ``dust cycle'' : the first term represents the
amount of dust removed from the ISM due to star formation, the second
takes into account dust pollution by stars while the third and fourth
terms represent dust accretion and destruction in the ISM,
respectively.  In this work, we used the same prescriptions used in
Gioannini et al. (2017).  Dust production are provided by taking into
account condensation efficiencies $\delta_i$
\footnote{The condensation efficiency ($\delta_i$) represents the fraction of an element i expelled from a star which goes into the dust phase of the ISM.}, as
provided by Piovan et al. (2011). 

 The dust yields  $\delta_i R_i(t)$ are not only
metallicity dependent but also depend on the mass of the progenitor
star.  In this work we consider as dust producers Type II SNe
($M_{\star}>8M_{\odot}$) and low-intermediate mass stars
($1.0M_{\odot}<M_{\star}<8.0M_{\odot}$).  Concerning dust accretion and
destruction we calculated the metallicity dependent time-scales for
these processes ($\tau_{accr}$ and $\tau_{destr}$) as described in Asano et
al. (2013).  For a more detailed explanation on dust prescriptions or
dust chemical evolution model we address the reader to Gioannini et
al. (2017).

\begin{figure}
\centering
\includegraphics[scale=0.75]{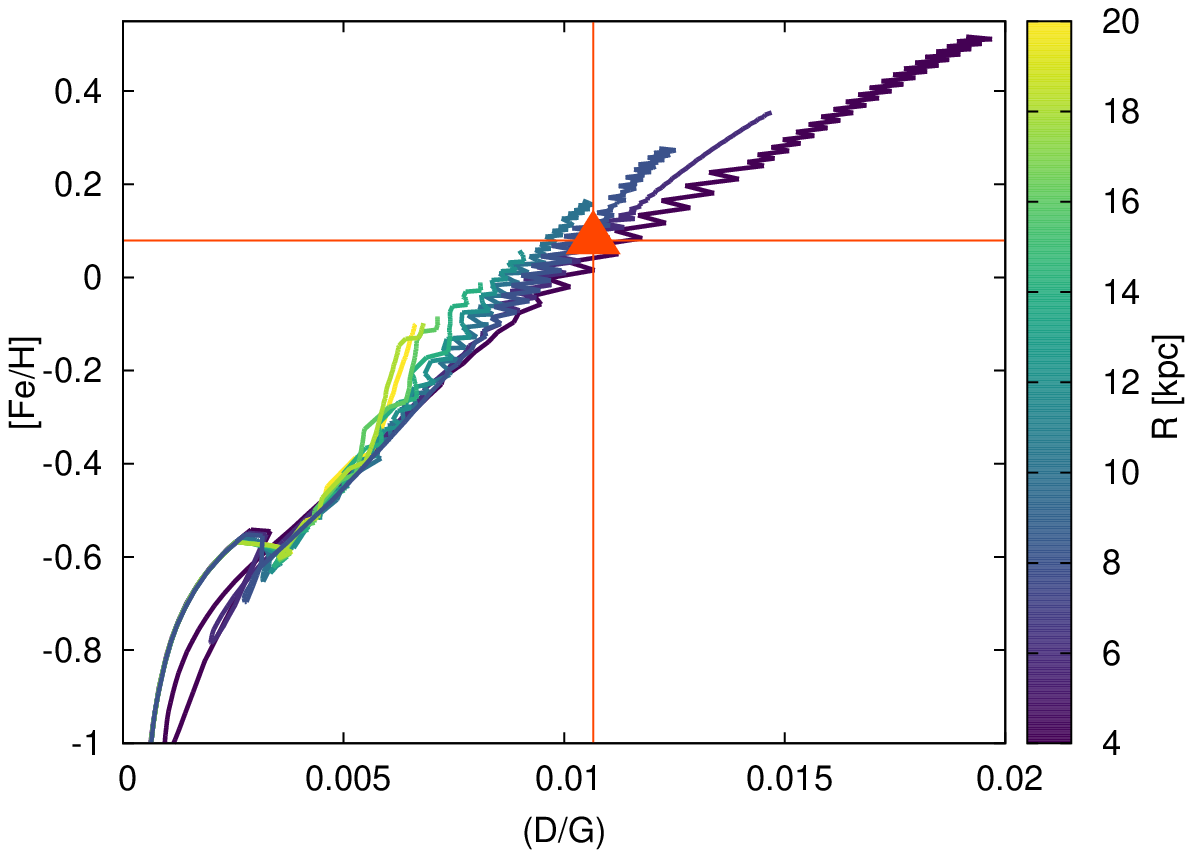}
\includegraphics[scale=0.7]{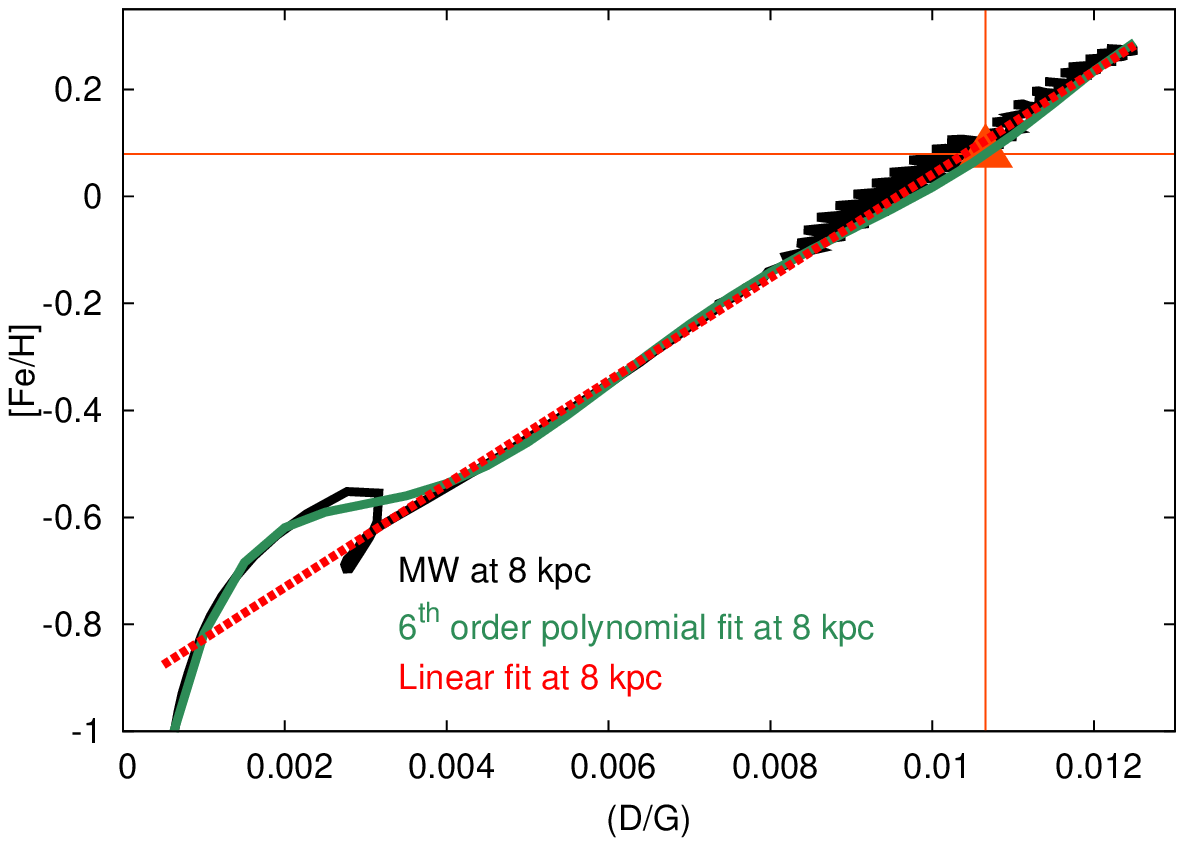}
 \caption{{\it Upper panel}: The evolution of the abundance ratio
   [Fe/H] as  the function of the dust-to-gas ratio\ $\left(
   \frac{D}{G}\right)$ predicted by our chemical evolution model of
   the Milky Way disc. As in Fig. \ref{SFR} the color code indicates
   different Galactocentric distances.  {\it Lower panel} With the
   black solid line we report the [Fe/H] ratio vs $\left(
   \frac{D}{G}\right)$ computed at 8 kpc (solar neighborhood) using
   the chemical evolution model of the Milky Way. With the green solid
   line we show the fit obtained by mean of a sixth order polynomial
   fit.  With the red dashed line we report the linear fit at 8 kpc.
   In both panels with the orange triangle we label the value of the
   [Fe/H] as a function of the dust-to-gas ratio $\left( \frac{D}{G} \right)$ for the model computed in the
   solar neighborhood at the Galactic time of 9.5 Gyr (i.e. model
   solar value).}
\label{DG}
\end{figure}

\section{Results}
 First, in this Section we present  the main results of our Milky Way chemical
 evolution model  in presence of dust.
 Moreover, the $P_{E/FGK,M}$ probabilities of finding Earth-like planets but not
 gas giant ones around FGK and M stars of Gaidos \& Mann (2014)
 computed with detailed chemical evolution models for the Galactic
 disc at different Galactocentric distances are shown. 
 We 
 express those probabilities in terms of the dust-to-gas ratio
 $\left(\frac{D}{G}\right)$ obtained by our ISM chemical evolution
 models.  Finally, we  present the maps of habitability of our
 Galaxy as functions of the galactic time and Galactocentric
 distances in terms of the total number of FGK and M stars which could
 host habitable Earth like planets and not gas giant planets. 

\begin{figure*}
\centering
\begin{tabular}{@{}p{0.45\linewidth}@{\quad}p{0.45\linewidth}@{}}
\subfigimg[width=\linewidth]{\hspace{ 5.7cm}  A)}{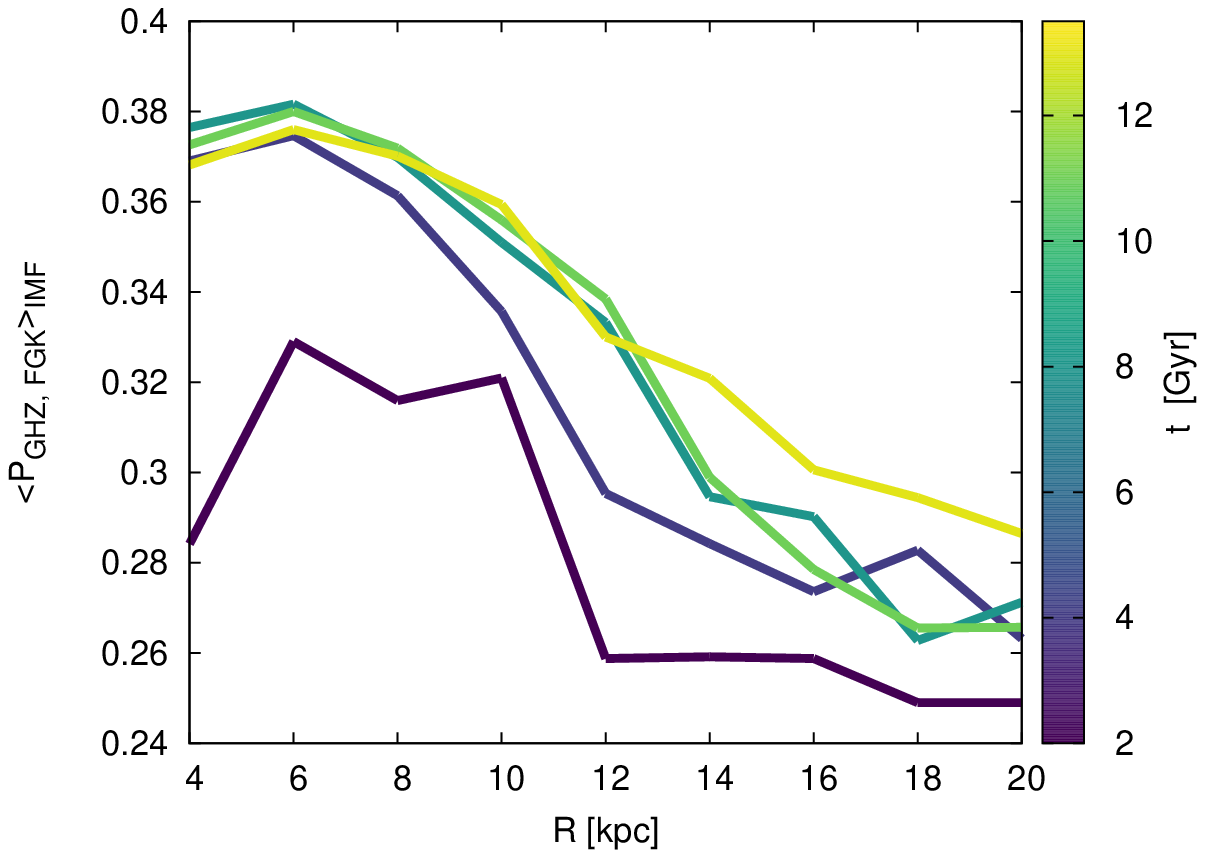} &
    \subfigimg[width=\linewidth]{ \hspace{ 5.6cm} B)}{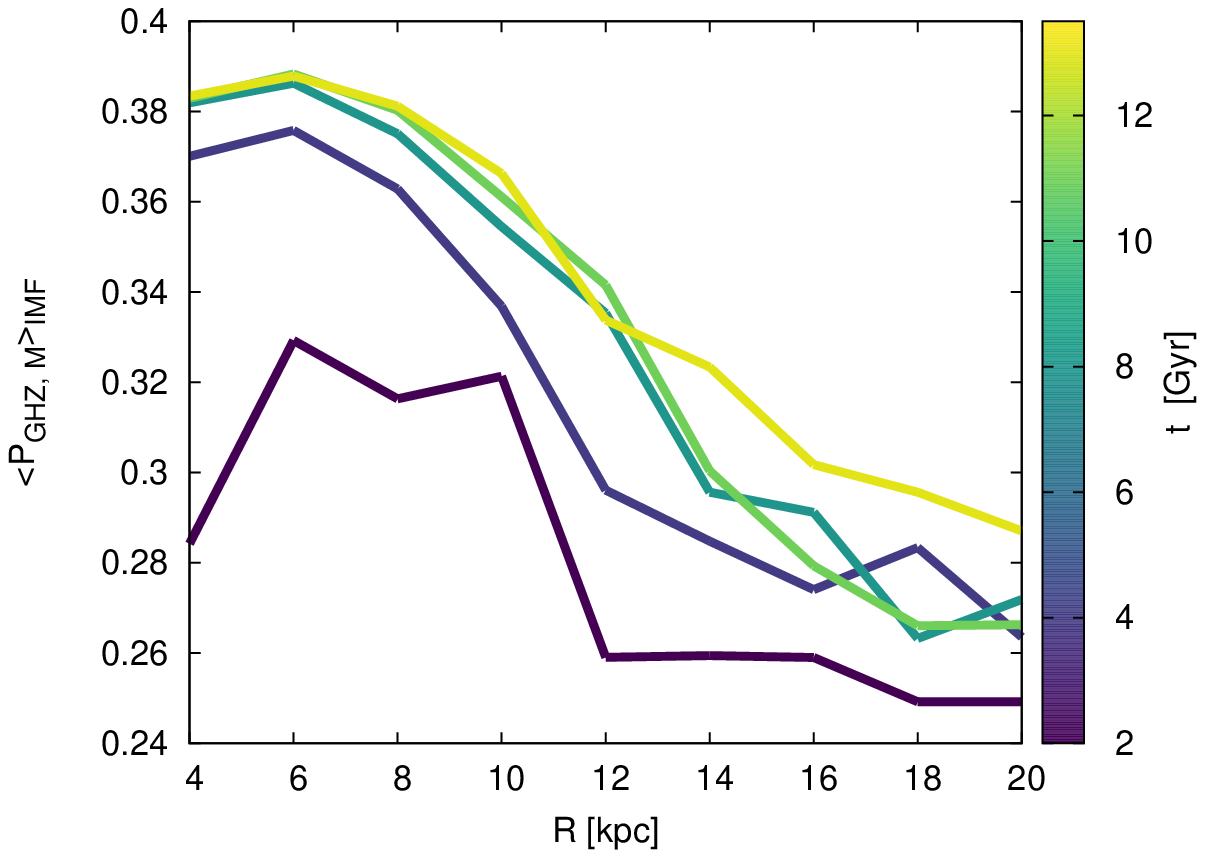} \\
    \subfigimg[width=\linewidth]{\hspace{ 5.6cm} C)}{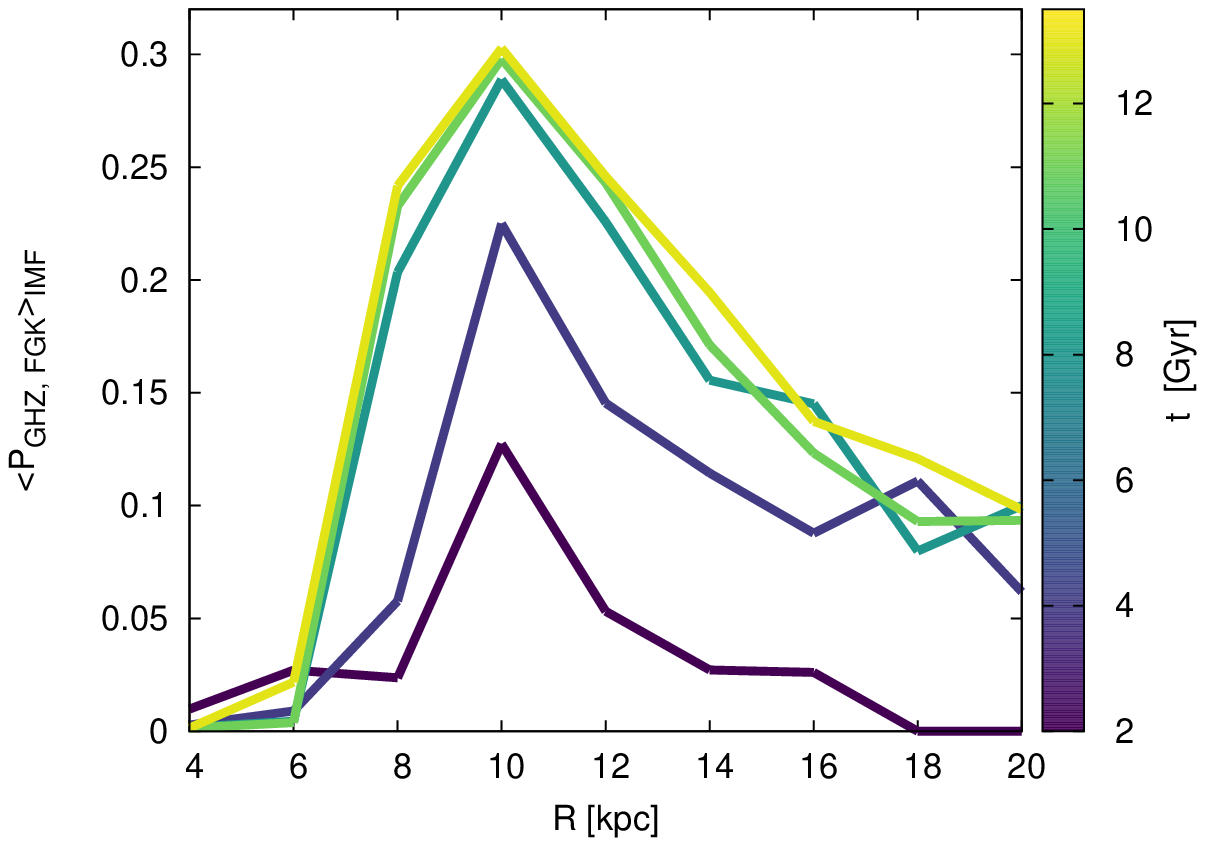} &
    \subfigimg[width=\linewidth]{\hspace{ 5.7cm} D)}{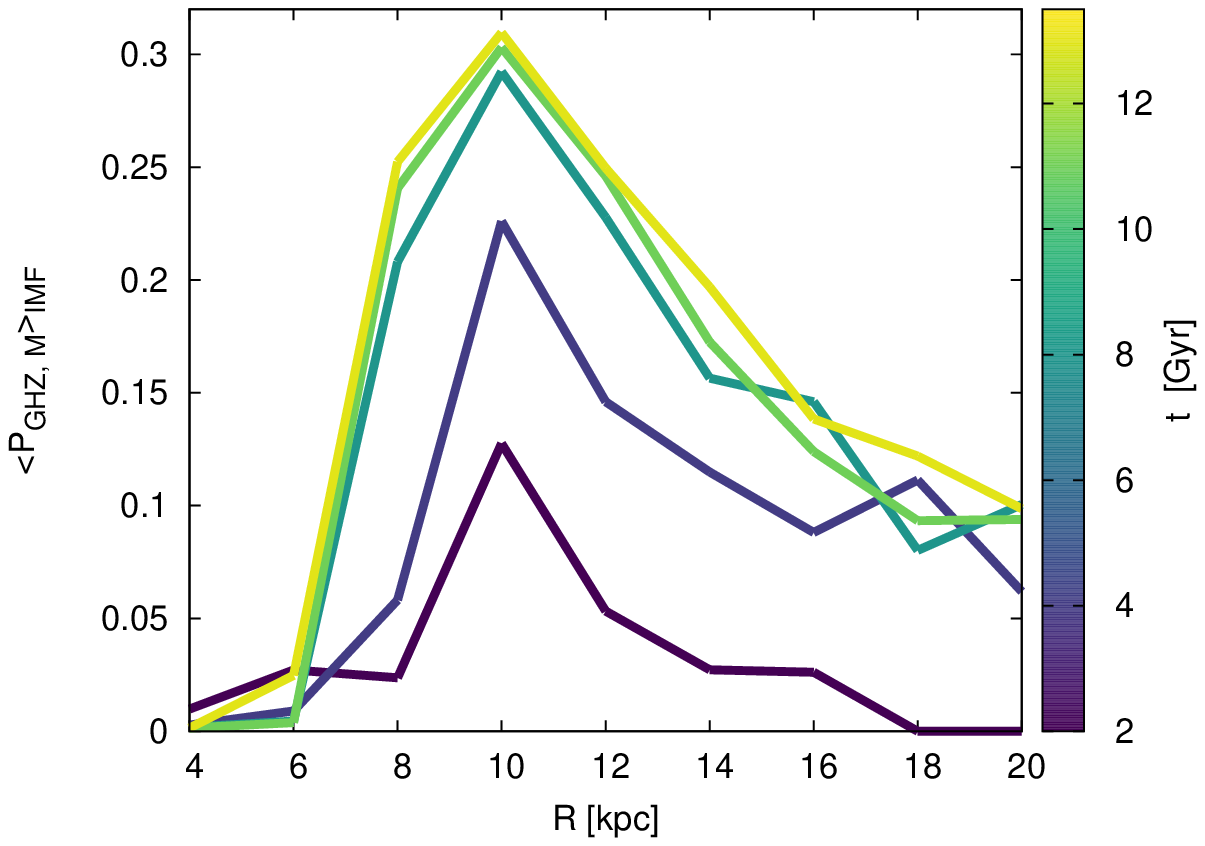}
  \end{tabular}
\caption{{\it Upper panels}: 
The probability $<$$P_{GHZ}$$>$$_{IMF}$ to find terrestrial habitable planets but
not gas giant around FGK stars ({\it Panel} A) and M stars ({\it
Panel} B) as the function of the Galactocentric radius. In this case
we do not consider the destructive effects from nearby SN
explosions. The color code indicates different Galactic times. {\it
Lower panels}: The probability $P_{GHZ}$ to find terrestrial habitable
planets but not gas giant around FGK stars ({\it Panel} C) and M stars
({\it Panel} D) as the function of the Galactocentric radius taking
into account the destructive effects from nearby SN explosions. The
color code indicates different Galactic times. } \label{NOSN} 
\end{figure*} 

\subsection{The Milky Way disc in presence of dust}
In Fig. \ref{SFR} we show the time evolution of the star formation
rate (SFR) (panel A), of [Fe/H] abundances (panel B), of SN rates
(panel C), and finally of the total dust (panel D) as functions of the
Galactocentric distance.  In panel A) we see the effect of the
inside-out formation on the SFR in the thin disc phase.  During the
halo-thick disc phase (up to 1 Gyr since the beginning of the star
formation) all the Galactocentric distances show the same star
formation history (for all radii we assume the same surface gas
density and same formation time-scales in the halo-thick disk phase,
for details see Section 3).

 In the inner regions the SFR in the thin disc phase is higher because
 of the larger gas density and shorter time-scales of gas accretion
 compared to the outer regions. Indeed, in the outer regions it is
 more evident the effect of the threshold in the gas density: the SFR
 goes to zero when the gas density is below the threshold.  We notice
 also that the in the halo-thick disc phase we have the same SFR
 history at all Galactocentric distances.

In panel B) it is shown the ``age-metallicity'' relation in terms of
[Fe/H] ratio vs Galactic time. It is clear also in this case the effects of the
inside-out formation: the inner regions exhibit a faster and more
efficient chemical enrichment with higher values of [Fe/H]. Actually,
at early times, in correspondence of the beginning of the second
infall of gas (thin disc phase) there is a  drop of the  [Fe/H] abundance
values. This drop is more evident in the inner regions of the Galactic
disc. This is due to the fact that in the inner regions the second
infall of primordial gas related to the thin disc phase is more
massive and on shorter time-scales, therefore the chemical abundances
are more diluted at the beginning of the thin disc phase in the inner
Galactic regions, compared to the external ones. 

 In panel C) of Fig. \ref{SFR} we present the total SN rates as functions of the
 Galactic time and Galactocentric distances. With the red line we
 label the limit adopted in this paper and in Paper I to
 take into account the destruction effects of SN explosions on the
 Galactic habitable zones modeling ($2 \times <RSN_{SV}>$). Above this
 SN rate limit we assume that there is zero probability to have life on
 a terrestrial planet.

Finally in panel D)  of Fig. \ref{SFR} we show the time evolution of the total surface
mass density of dust at different Galactocentric radii.  Dust
production by stars is the main source of dust in the early phases of
the Milky Way evolution.  For this reason, in the inner regions, the
dust amount is higher because of the large production by Type II SNe,
during the initial burst of star formation, as visible in panel A).
On the other hand, dust accretion becomes important at later epochs,
and the dust mass tends to increase at all Galactocentric distances.
The observed oscillation of the model occurs when the rates of dust
accretion and dust destruction are comparable.  In fact, in this case
there is a gain of the total mass surface density of dust provoked by
the dust growth, rapidly followed by a decreasing due to the dust
destruction rate, which exceeds dust accretion.  This turnover between
those processes occurs especially in the quiescent phases of the
Galactic evolution.

\subsection{The computed probabilities $P_{GGP/FGK,M}$ with the two infall chemical evolution model}
To compute the probabilities $P_{GGP/FGK,M}$ presented in
eqs. (\ref{eqgiantF}) and (\ref{eqgiantM}) with our chemical evolution
model we consider the weighted values on the IMF using the following
expressions:
\begin{displaymath}
  <P_{GGP/FGK}(R,t)>_{IMF}=
\end{displaymath}

\begin{equation}
  0.07 \times 10^{1.8 \mbox{[Fe/H]}(R,t)_{model}} \left( <\frac{M_{\star, FGK}}{M_{\odot}}>_{IMF}\right),
\end{equation}
for the FGK stars, and
\begin{displaymath}
  <P_{GGP/M}(R,t)>_{IMF}=
\end{displaymath}

\begin{equation}
  0.07 \times 10^{1.06  \mbox{[Fe/H]}(R,t)_{model}} \left( <\frac{M_{\star, M}}{M_{\odot}}>_{IMF}\right),
\end{equation}
 for M stars.

The [Fe/H]$(R,t)_{model}$ quantity   is the  computed iron abundance with our chemical evolution model adopting  the Scalo (1986) IMF at the 
Galactic time $t$ and Galactocentric distance $R$.

Therefore, to compute  $<P_{GGP/M}>_{IMF}$ and $<$$P_{GGP/FGK}$$>$$_{IMF}$
quantities we have only to know the weighted stellar mass on the IMF in the
mass range of M stars and FGK stars, respectively.  
The weighted
stellar masses on the   Scalo (1986) IMF are:
\begin{equation}
 <\frac{M_{\star, FGK}}{M_{\odot}}>_{IMF}=\frac{\int_{0.45 \, M_{\odot}} ^{1.4 \,  M_{\odot}} m^{-1.35} dm} {\int_{0.45 \,  M_{\odot}} ^{1.4 \,  M_{\odot}} m^{-2.35} dm}=0.72584
\end{equation}
and
\begin{equation}
 <\frac{M_{\star, M}}{M_{\odot}}>_{IMF}=\frac{\int_{0.08 \, M_{\odot}} ^{0.45 \,  M_{\odot}} m^{-1.35} dm} {\int_{0.08 \,  M_{\odot}} ^{0.45 \,  M_{\odot}} m^{-2.35} dm}=0.15504.
\end{equation}
In Fig. \ref{giant} we show the evolution of $<P_{GGP/FGK}>_{IMF}$ and
$<P_{GGP/M}>_{IMF}$ probabilities as the function of the [Fe/H]
abundance ratio computed at different Galactocentric distances using
our chemical evolution models. Because of the inside-out formation, the
inner regions exhibit a faster and more efficient chemical
enrichment. In fact, the model computed at 4 kpc reaches, at the
present time, [Fe/H] value of 0.55 dex, instead of at 20 kpc the
maximum [Fe/H] is equal to -0.1 dex. We see that the two probabilities become
to be substantially different for over-solar values in the inner
regions.  For instance, at the present time, i.e. at the maximum
values of the [Fe/H] abundance  in Panel C of
Fig. \ref{SFR}, at 4 kpc the probabilities $<P_{GGP/FGK}>_{IMF}$ and
$<P_{GGP/M}>_{IMF}$ show the values of 0.43 and  0.04, respectively.

\subsection{The dust-to-gas ratio $\left( \frac{D}{G} \right)$}
In this Subsection we provide a useful theoretical tool to set
the proper initial conditions for the formation of protoplanetary discs.
As underlined in Section 1,  while dust grains of $\mu$m are
directly observed in protoplanetary discs, on the other hand, the
amount of gas mass  is set starting
from the one of the dust and by assuming a value for the dust-to-gas
ratio. Unfortunately, this practice has several 
uncertainties (Williams \& Best 2014).
In this subsection, we connect the evolution of the dust to gas ratios at 
different Galactocentric distances with the chemical enrichment expressed
in terms of [Fe/H]. Because of the well known ``age-metallicity'' relation 
 reported in the Panel C) of Fig. \ref{SFR},  the dust-to-gas ratio ($\frac{D}{G}$) vs  [Fe/H] abundance ratio relation  can be seen as a time evolution  for the dust-to-gas ratio  $(D/G)(t)$. 

 In the upper Panel of Fig.~\ref{DG} is presented the evolution of the
 dust-to-gas ratio ($\frac{D}{G}$) as a function of [Fe/H] at
 different Galactocentric distances.  As expected, higher
 metallicities are reached in the inner radii, where the star
 formation is higher.  The dust-to-gas ratio ($\frac{D}{G}$) 
increases in time for two reasons: the first is that dust
 production in star forming regions is high, especially from Type II
 SNe, while the second is related to dust accretion.  Dust accretion
 is a very important process occurring in the ISM and it becomes the
 most important one as the critical metallicity is
 reached\footnote{The critical metallicity is the metallicity at which
   the contribution of dust accretion overtakes the dust production
   from stars (Asano et al. 2013).}.

As dust grains are formed by metals, dust accretion becomes more
efficient as the metallicity in the ISM increases. For this reason at
high values of [Fe/H], we found higher values of dust-to-gas ratio.
The relation between the [Fe/H] and the dust-to-gas ratio is important
because it can provide the probability of planet
formation depending on the amount of dust in the ISM and, on the other
hand, provides an estimate of the dust-to-gas ratio in the ISM during the
formation of a protoplanetary disc.
 
The solar dust-to-gas ratio $\left( \frac{D}{G} \right) _{\odot}$ value predicted by our model (model value computed in the solar neighborhood at 9.5 Gyr) is 0.01066.

In the lower panel of Fig.~\ref{DG} we show the $\frac{D}{G}$ as a
function of the [Fe/H] values only for the shell centered at 8 kpc and
2 kpc wide (the solar neighborhood).  In the same plot the sixth
degree polynomial fit which follows exactly the [Fe/H] vs
$\frac{D}{G}$ in the range of [Fe/H] between -1 dex and 0.5 dex is presented. The expression of this fit  is the
following one:

\begin{equation}
\mbox{[Fe/H]}=\sum_{n=0}^{6} \alpha_n \left(\frac{D}{G}\right)^n 
\end{equation}
All the coefficients $\alpha_n$ and $n$ are reported in the footnote \footnote{\mbox{[Fe/H]}=$-1.48 +
 1.05 \,10^4 \left(\frac{D}{G}\right) -
 4.91 \,10^5 \left(\frac{D}{G}\right)^2 + 1.14
 10^8 \left(\frac{D}{G}\right) ^3 -
 1.32 \,10^{10} \left(\frac{D}{G}\right)^4 +
 7.57 \,10^{11} \left(\frac{D}{G}\right)^5 -
 1.69 \,10^{13} \left(\frac{D}{G}\right)^6$}.

We found  that for [Fe/H] values higher than -0.6 dex a
linear fit is able to reproduce pretty well the computed [Fe/H] vs $\left(
\frac{D}{G} \right)$ relation in the solar vicinity.

The equation of the linear fit reported in the lower panel of Fig.~\ref{DG} is:
\begin{equation}
\mbox{[Fe/H]}=96.49 \left( \frac{D}{G} \right) -0.92.
\end{equation}

This relation is important to connect the dust-to-gas ratio with the
[Fe/H] abundance. If we  combine it  with eq.  (\ref{ter})
we obtain  the probability of having  terrestrial planets  but not gas giant
ones depending on the amount of dust in the ISM.

\subsection{The $P_{GHZ}$ values around  FGK and M stars}

In the upper panels of Fig. \ref{NOSN} (A and B) we show the evolution
in time of the $P_{GHZ}$ values for FGK and M stars as  functions of
the Galactocentric distance in the case where SN destruction effects
are not taken into account.

We notice that the $<P_{GHZ/M}>_{IMF}$ and $<P_{GHZ/FGK}>_{IMF}$
probabilities are identical at large Galactocentric distances. This is
due to the fact that, as shown in Fig. \ref{pe}, the $<P_{E/FGK}>_{IMF}$ and
$<P_{E/M}>_{IMF}$ probabilities are similar for sub-solar values of
[Fe/H]. The chemical evolution in the outer parts of the Galaxy,
because of the inside-out formation, is slow and with longer time
scales. The maximum values of [Fe/H] are smaller than in the inner region
and sub-solar (see the ``age-metallicity'' relation reported in panel
B of Fig. \ref{SFR}).

Moreover, in the inner regions $<P_{GHZ,M}>_{IMF}$ and
$<P_{GHZ,FGK}>_{IMF}$ probabilities become to be different only for
Galactic times larger than 8 Gyr. As expected the higher 
probabilities are related to the M stars.  Even if the $<P_{E/FGK}>_{IMF}$ and
$<P_{E/M}>_{IMF}$ probabilities are substantially different for [Fe/H]$>$ 0.2
(see Fig. \ref{pe}), the two associated $P_{GHZ}$ probabilities are similar.

 This is due to the definition of $P_{GHZ}(t)$: at each Galactic time
 the SFR $\times$ $P_{E}$ quantity is integrated from 0 to $t$. In
 other words, we are weighting the $P_{E}$ quantity on the SFR. From
 panel C) of Fig. \ref{SFR} it is clear that the peak of the SFR in
 the inner regions (annular region between 3 and 7 kpc) is around 3
 Gyr.  From the ``age-metallicity'' relation reported in panel B) of
 Fig. \ref{SFR}, we
 derive that at this age the mean Galactic [Fe/H] is -0.5 dex.  At
 this metallicity, as stated above,  the $P_{E.FGK}$ and $P_{E.M}$
 values are almost the same.  This is the reason why the two $P_{GHZ}$
 probabilities are similar even at late time in the inner regions.

In the lower panels of Fig.\ref{NOSN} (C and D) we show that, the
$P_{GHZ}$ probabilities for FGK and M stars as functions of the
Galactocentric distances and time, when the SN destruction effects
have been taken into account, are almost identical.  As found in Paper
I the Galactocentric distance with the highest probability that a star
(FGK or M type) hosts a terrestrial planet but not gas giant ones is
10 kpc.

\begin{figure}
          \centering \includegraphics[scale=0.499,
     angle=-90]{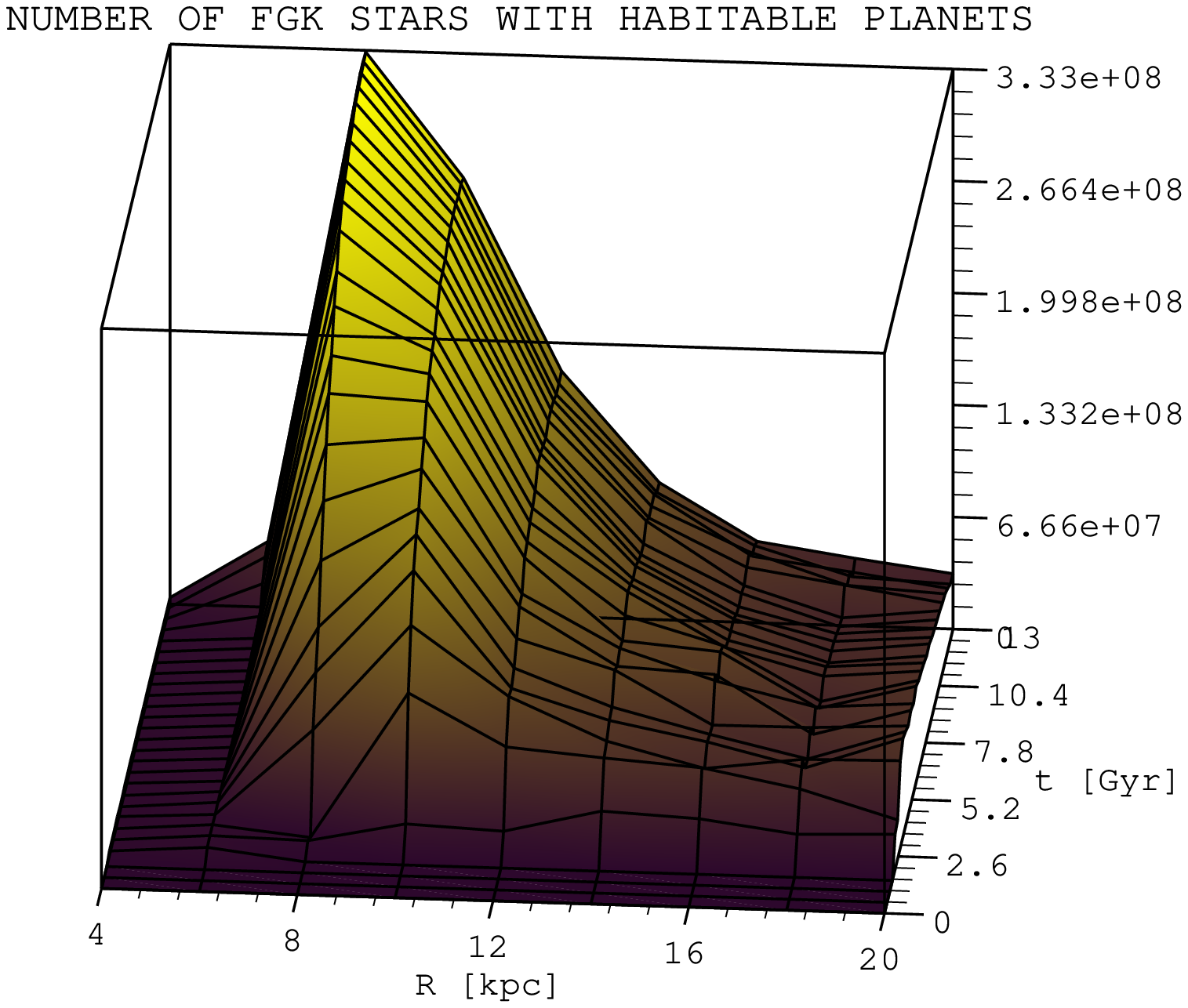} \includegraphics[scale=0.499,
     angle=-90]{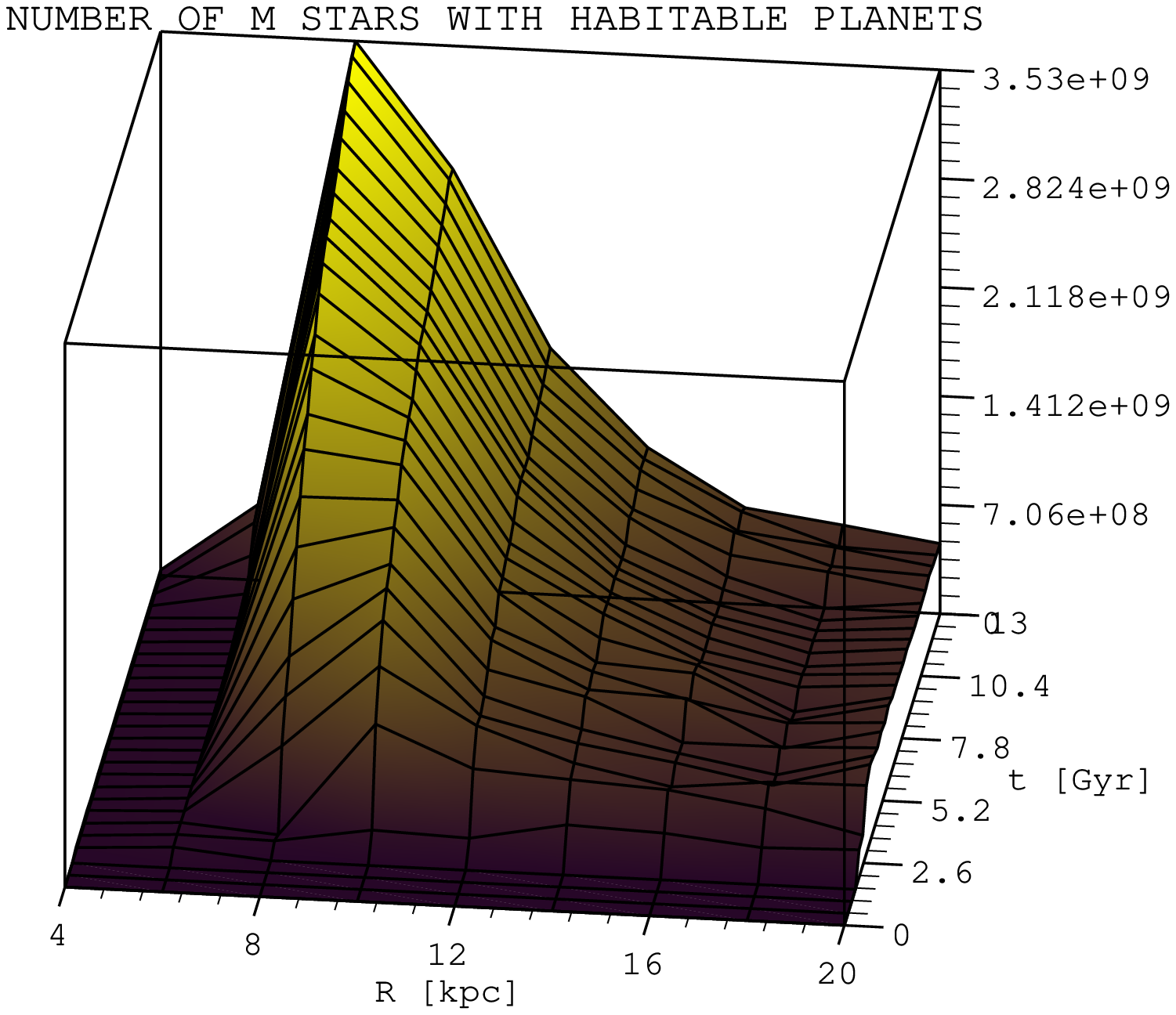} \caption{The total number of FGK
     stars ({\it Upper panel}) and M stars ({\it Lower  panel}) having habitable terrestrial planets but not gas giant ones as functions of the Galactocentric distance and the  Galactic time (the $N_{\star \, life}$ quantity in eq. \ref{star}) where  SN destructive effects are taken into account. The number of stars  are computed
     within concentric rings, 2 kpc wide.}  \label{maps}
\end{figure}

\subsection{THE GHZ maps for FGK and M stars}

In order to recover the total number of stars at the Galactic time $t$
and Galactocentric distance $R$ hosting habitable planets, it is
required the total number of stars created up to time $t$ at the
Galactocentric distance $R$ (the $N_{\star tot}$ quantity in
eq. \ref{star}).

Our chemical evolution model very well reproduces the observed total
local stellar surface mass density of 35 $\pm$ 5 M$_{\odot}$ pc$^{-2}$
(Gilmore et al. 1989, Spitoni et al. 2015).  In fact, our predicted
value for the total surface mass density of stars is 35.039 M$_{\odot}$
pc$^{-2}$. Morever, we find that this value concerning only to M stars is
24.923 M$_{\odot}$ pc$^{-2}$ and the one for FGK stars is 9.579
M$_{\odot}$ pc$^{-2}$.

In Fig. \ref{maps} we show the number of FGK stars (upper panel) and M
stars (lower panel) hosting habitable terrestrial planets but not gas
giant planets as functions of the Galactic time and Galactocentric
radius (the quantity $N_{\star \, life}$ of eq. \ref{star}).

We
notice that for both FGK and M stars, the GHZ maps, in terms of the
total number of stars hosting planets with life, peaks at 8 kpc.  On
the other hand, as we have seen above the maximum fraction of stars
which can host habitable terrestrial planets peaks at 10 kpc (see
panels C and D of Fig. \ref{NOSN}).

The reason why the GHZ peaks at galactocentric distances smaller than
in the case when it is expressed in terms of fraction of stars is the
following one: in the external regions the number of stars formed at
any time is smaller than in the inner regions because of the 
the smaller SFR.  This is in agreement with
the results of Prantzos (2008) and Paper I.

We see that, at the present time, in the solar neighborhood the number
$N_{\star, M, life}$/$N_{\star, FGK, life}$=10.60.  
This ratio is
consistent with the IMF we adopt in our model.  In fact, the ratio
between the fraction  of M stars over FGK stars (by number) in a
newborn population adopting a Scalo IMF is:
\begin{equation}
\left( \frac{M_{number}}{FGK_{number}}\right)_{\mbox{Scalo IMF}}=\frac{\int_{0.08 \, M_{\odot}} ^{0.45 \,  M_{\odot}} m^{-2.35} dm} {\int_{0.45 \,  M_{\odot}} ^{1.4 \,  M_{\odot}} m^{-2.35} dm}=11.85.
\end{equation}

Finally, the predicted local surface mass density of M stars hosting
habitable planets predicted by our model is 5.446 M$_{\odot}$
pc$^{-2}$, and the value for FGK stars is 2.40 M$_{\odot}$ pc$^{-2}$.

\section{Conclusions}
In this work we investigated the Galactic habitable zone of the Milky
Way adopting the most updated prescriptions for the probabilities of
finding terrestrial planets and gas giant planets around FGK and M
stars.  To do that we adopted a chemical evolution model for the Milky
Way which follows the evolution of the chemical abundances both in the gas
and dust.

The main results can be summarized as follow:
\begin{itemize}
\item Adopting the Scalo (1986) IMF the probabilities of finding gas
  giant planets around FGK and M stars computed with the two-infall
  chemical evolution model of Paper I begin to be
  different for supersolar values of [Fe/H]. In particular,
  substantial differences are present in the annular region centred at
  4 kpc from the Galactic centre.
\item We provide for the first time a sixth order polynomial fit (and
  a linear one but more approximated) for the relation found in the
  chemical evolution model in the solar neighborhood between the
  [Fe/H] abundances and the dust-to-gas ratio $\left(\frac{D}{G}
  \right)$.  With this relation it is possible to express the Gaidos
  \& Mann (2014) probabilities of finding gaseous giant planets around
  FGK or M stars in terms of the gas to dust ratio $\frac{D}{G}$.

\item We provide a useful theoretical tool to set the proper initial
  condition for the formation of protoplanetary disc connecting the
  evolution of the dust-to-gas ratios at different Galactocentric
  distances with the chemical enrichment expressed in terms of [Fe/H].

\item The probabilities that a FGK or M star could host  habitable planets are
 roughly 
identical. Slightly differences arise only at Galactic times larger
 than 9 Gyr where the probability of finding gas giant planets around FGK 
becomes substantially different from the one associated to M stars.
\item As found by Paper I, adopting  the same prescriptions for
  the destructive effect from close-by SN explosions, the larger  number
  of FGK and M stars with habitable planets are in the solar neighborhood.

\item At the present time the total number of M stars with habitable
  terrestrial planets without gas giant ones are $\simeq$ 10 times the
  number of FGK stars. This result is consistent with the Scalo (1986)
  IMF adopted here.

\end{itemize} 

The Gaia mission with its global astrometry, will be crucial for the
study of exoplanets. We recall the relation used in this work between
the frequency of gas giant planets and the metallicity of the host
star was obtained by means of the Doppler effect with the method of
radial velocity, and it will be possible with Gaia to test whether
this result is an observational bias or is related to real physical
processes.  It is estimated that Gaia will be able to find out up to
$10^4$ giant planets in the solar vicinity with its global astrometry
with distances spanning the range between 0.5 and 4.5 AU from the host
star.

\section*{Acknowledgments}
 We thank the anonymous referee for the suggestions that
improved the paper. The work was supported by PRIN MIUR 2010-2011, project ``The Chemical
and dynamical Evolution of the Milky Way and Local Group Galaxies'',
prot. 2010LY5N2T.

\end{document}